\documentclass[epj,referee]{svjour}
\usepackage{graphics}
\begin{document}
\title{Fluid and plastic flow dynamics
of the critical state for a strongly pinned 2D superconductor}
\subtitle{} \titlerunning{Fluid and plastic flow dynamics ...}
\author{D. Monier \and L. Fruchter}%
\institute{Laboratoire de Physique des Solides, C.N.R.S.
Université Paris-Sud, 91405 Orsay cedex, France }
\date{Received: date / Revised version: date}
%
\abstract{ We present simulations of the dynamic critical state
for a 2D superconductor with strong pinning centers,
corresponding to a matching field twice the applied magnetic
field. The sharp crossover between the plastic regime, at low
current density and temperature, and the fluid flow regime for
flux motion is characterized by the activation energy for flux
motion and the transverse diffusion of the vortices trajectory.
} 

\maketitle
\section{Introduction}
\label{intro} Flux pinning in type II superconductors is a well
documented example for the general problem of an elastic periodic
medium interacting with disorder. The study of the dynamics of the
pinned flux line lattice, under the action of an external current
or by the virtue of a time varying external magnetic field, is
one way to apprehend the problem of flux line pinning. Well known
examples are transport I-V characteristics and magnetization
relaxation experiments. Such experiments are usually interpreted
using simple models, assuming a uniform critical current density
and a single, current dependent, activation energy at non zero
temperature. However, numerical
simulations\cite{jensen90,nori96,reichhardt96,dominguez99},
experiments\cite{matsuda96,marchevsky97} and
theory\cite{giam96,led98}, have shown that a driven flux lattice
can be a rather complex object, remarkably different from the
simple pictures used in the interpretation of the experiments. In
particular, the pinning of a two-dimensional flux lattice by
strong pinning centers lends itself easily to numerical
simulations and has been the subject of several studies. Such a
system is also a good approximation for highly anisotropic
superconductors irradiated by heavy ions, such as the high Tc
oxides, which are now well documented. In this paper, we present
numerical simulations of this strong pinning situation, at non
zero temperature and for conditions similar to those encountered
in experimental magnetization experiments.

\section{Simulation}

We simulate a slab submitted on one side to an external magnetic
field, applied parallel to its boundary, while, on the opposite
side, the external magnetic field is assumed to be zero (vortices
are removed as they cross this boundary). Periodic conditions are
used, so that the effect of the finite dimensions is limited. Flux
lines are assumed rigid rods. This correctly modelizes a layered
superconductor with decoupled layers in the critical state, for
which the local magnetic induction decreases from the applied
field value on the external sides to a lower value at the center
of the sample (Fig. \ref{sketch}).

Strong pinning centers are randomly distributed in the sample.
The density of the pinning sites is $n=B_{\Phi}/\Phi_{0}$, with
$\Phi_{0}$ the flux quantum and $B_{\Phi}$ the 'matching field'
for which an equilibrium flux line lattice shows the same density
of flux lines. A region of width $\lambda$ from the side where
vortices are introduced is kept free of pinning sites, in order
to allow for the initial formation of a regular hexagonal lattice
in this region. The pinning sites are assumed normal cylinders
parallel to the applied field, with radius $c_0$. The situation
where, at low temperature, the vortex core radius $\xi(T)$ is
smaller than $c_0$ and pinning is due essentially to the
reduction of the core energy when the line sits on the pin is
considered. The force exerted by a pin at a distance $r$ to the
line is given by:
\begin{equation}
  f_{p}(r) = \left\{\begin{array}{l}
    \sigma\varepsilon_{o} r/r_{0} \xi\;\;\mbox{for}\;r\leq r_{0}\;,\\
    0\;\;\mbox{for}\;r>r_{0}\;.\\
  \end{array}\right.
\end{equation}
where $\varepsilon_{0} = \left(\Phi_{0}/4\pi\lambda\right)^{2}$
is the line energy, $r_{0}=c_{0}+\xi/2$ and $\sigma \leq1$.

The force per unit length exerted by a vortex at a distance $r$
to the line is:
\begin{equation}
f_{vv}(r)=\left(\Phi_0^{2}/8\pi^{2}\lambda^{3}\right)
K_{1}\left(r/\lambda\right)
\end{equation}
where $K_{1}$ is a Bessel function. This is a good approximation
strictly only in the case of vortex lines (rods) and for 2D
vortices a logarithmic interaction should be used. In the present
case, the more rapid decrease of the Bessel function allows us to
cut the interaction between vortices at a distance $5\lambda$ and
save computation time.

The external magnetic field $B_{0}$ is simulated by an extra force
$f_{B_{0}}$  acting on each vortex, perpendicular to the external
side of the sample. The force acting on a vortex at a distance
$x$ from the boundary is the one imposed by a semi-infinite
vortex lattice at a distance $a_{0}+x$, where
$a_{0}=\left(\Phi_{0}/B\right)^{1/2}$ is the flux lattice spacing
at the equilibrium.

The finite temperature is simulated in a way similar to the one
described in ref. \cite{brass89} by adding a stochastic velocity
to the particles:
\begin{equation}
{\bf v_{T}} = \left(2\;k_{B}\;T\;\tau/3\Delta^{2}\right)^{1/2}\;
\left(
\begin{array}{c}
\gamma_{x}\\
\gamma_{y}\
\end{array}
\right) \Theta(\Delta t/\tau - q)
\end{equation}
where $1/\tau$ is the average frequency for the thermal
perturbation, $\gamma_{x,y}$ are random numbers from a Gaussian
distribution of width 1, $q$ is a random number from a uniform
distribution between 0 and 1, and $\Theta$ is the Heaviside
function. We have used a single set of parameters $\tau$ and
$\Delta t$ for all temperatures with $\tau \gg \Delta t$. The
accuracy for the temperature simulated in this way was checked by
computing the life time of a single vortex in a potential well
similar to the ones defined above. The life time as a function of
the well depth was found to match accurately the one given by the
Arrhenius-Kramers relation for an overdamped particle, in the
range of temperature studied here. In this overdamped regime,
results are independent of the viscosity when expressed in the
time unit $t/\eta$, the escape frequency being inversely
proportional to $\eta$.

Finally, the displacement $\Delta {\bf r_{i}}$ of vortex $i$
during time step $\Delta t$ is computed according to the
diffusive equation:
\begin{eqnarray}
&&\eta\;\Delta{\bf r_{i}}/\Delta t =\nonumber\\
&&\sum_{j\neq i}\;{\bf f_{vv}} (r_{ij}) +\sum_{j} \; {\bf f_{p}}
(r_{ij})+ {\bf f_{B_{0}}}(x_{i}/a_{0})+\eta\;\bf {v_{T}}
\label{diffusiveequation}
\end{eqnarray}
where the first sum is over all vortices at a distance less than
$5\lambda$ and the second one is over all pins ($\eta$ is the
viscosity per unit length). The time step is chosen so that the
vortex displacement during one step is always much smaller than
the pinning characteristic length, $r_{0}$. At non zero
temperature, vortices tend to creep through the sample from the
boundary submitted to the external field, where vortices are fed
whenever the magnetic pressure is strong enough to allow for
their penetration, to the one with zero applied field. As a
consequence, a steady thermally activated flux flow is set after
some time, characterized by a linear flux density profile and a
steady vortex current density, $\mathcal{F}$, that flows through
the sample.

\section{Results}

In the rest, we use the following material parameters,
appropriate for a high temperature superconductor:
$\lambda=1400\;$\AA, $\xi=18\;$\AA, $c_{0}=35\;$\AA $\;$ and
$\sigma=0.1$. The samples held typically 5000 pinning sites and
1000 vortices in the steady state. The matching field,
$B_{\Phi}=5000\;G$ and the external field, $B_{0}=2500\;G$, were
the same for all the simulations.

In a way similar to what is done in the analysis of real
relaxation experiments, we use a one-dimensional model with a
single activation energy to analyze the vortex current
density\cite{beek92}. According to this widely used model,
vortices thermally hop over barriers disposed along a 1-D axis
with height $U(j)$ (where $j = (4\pi)^{-1}\partial B/\partial x$
is the current density) and the vortex current density is ${\cal
F} =  j\;B\;\phi_{0}\;\eta^{- 1}\exp(-U(j)/k_{B}T)$. This model
clearly misses some of the fundamental features of the real
experiments. As pointed in ref. \cite{beek92}, it is valid only
in the case of almost uniform vortex and current densities, when
physical quantities may be replaced by their average over several
vortex spacing. In particular, it assumes that the screening
current flows in our case along straight lines perpendicular to
the vortex flow. As can be seen in Fig.~\ref{lattices}A, this is
not so and the current tends to flow along some curved, branched
paths that reveal the highly inhomogeneous stresses in the pinned
flux lattice. There has been attempts to account for the spatial
heterogeneity of the real systems, using a distribution of
activation energies\cite{griessen91,hoekstra99}. However, the
method assumes parallel, independent relaxation channels, which
is still a very crude assumption. As for real magnetization
experiments analysis, we obtain the average screening current
density from the total magnetization of the sample. In order to
investigate different flux gradient profiles, different sample
geometries were used with the length of the cylinder between
$20\lambda$ and $100\lambda$ and diameters between $10\lambda$
and $300\lambda$. The experiments probed different $T$ and $j$
values, which are located on a smooth ($T$, $j$) trajectory shown
in Fig.\ref{tju}. This is similar to relaxation rate measurements
in superconductors, for which both the temperature and the
screening current are varied  - the latter parameter being
implicitly determined by the temperature and the time window
explored in these experiments.

 The activation energy obtained from the simulations is shown
in Fig.~\ref{tju}, where energy is normalized to the single
vortex pinning site energy, $U_{0}=\sigma \varepsilon_{0} r_{0} /
\xi$, and critical current density to the single vortex critical
current, $j_{0}=\sigma \varepsilon_{0} / \xi \phi_{0}$. It is
shown as a 3D plot, as the computation of the energy with one of
these two parameters fixed would imply prohibitive computation
time as temperature or current decreases. The same limitation is
also encountered in real relaxation experiments, although in this
case the temperature variations are usually disregarded and the
data plotted as $U(j)$. The reason for this is that Maley's
procedure\cite{maley90} allows to show that, in most cases, the
activation energy obtained from relaxation experiments is
temperature-independent. We notice, however, that this is no
longer true at low temperature, which has been attributed to the
occurrence of the quantum tunneling regime\cite{monier98}.

A clear upturn can be seen in Fig.~\ref{tju}, where the activation
energy increases strongly with decreasing current below $j/j_{0}
\simeq 0.05$, $k_{B}T/\Delta \simeq 0.25$. This upturn
corresponds to the crossover between a plastic to a fluid flow
regime for flux motion. This can be seen, first, from a direct
observation of the vortices trajectories: in the plastic regime,
there exists a time scale at which some vortices move over
distances larger than $a_{0}$, while others remain immobile on
that scale and constitute some kind of 'pinned islands'
(Ref.\cite{nori96}). It is important to stress that these islands
depin at a larger time scale for non zero temperature, so that
only a dynamical definition of these domains may be given in our
case. No such characteristic time may be found in the fluid flow
regime where, depending on the time scale considered, nearly all
or none of the vortices move over this distance
(Fig.~\ref{lattices}A,B). A more quantitative argument in favor
of two clearly distinct regimes is given by the examination of
the diffusion of a vortex trajectory, transverse to the average
flux flow (Fig.~\ref{diffusion} and ~\ref{orderparameter}). In
the fluid flow regime, the transverse displacement on the scale
of the equilibrium flux lattice parameter, $\delta=a_{0}^{-1}
\langle [y(0)-y(a_{0})]^{2} \rangle^{1/2}$ where the average is
performed along the vortex trajectory, is significantly smaller
than 1, whereas it increases with decreasing temperature and
increasing current density up to a value $\simeq 0.6-0.65$ in the
plastic regime (Fig.~\ref{orderparameter}). This limiting value
may be understood by considering that the most disordered
trajectory allowed in the flux lattice is the one where a vortex
hops randomly from one interstitial site to a neighbor site. A
consequence of disorder in the plastic regime is that the
examination of the flux pattern hardly reveals the orientation of
the average flux flow (Fig.~\ref{lattices}C, making abstraction
of the region close to the zero field boundary of the sample,
where side effect is visible), as compared to the fluid flow
regime (Fig.~\ref{lattices}B). As can be seen in
Fig.~\ref{orderparameter}, the increase of the transverse
displacement along the $(T,j)$ trajectory shown in Fig.~\ref{tju}
is rather abrupt, suggesting that the trajectory might cross a
transition line rather than a crossover line. If this is so, an
order parameter may be elaborated from $\delta$, after subtraction
of the high temperature baseline and adequate normalization,
which is unity in the ordered regime and tends towards zero in
the opposite limit, thus delimiting two distinct, nonequilibrium,
phases. The static topological orders of the two phases which
would be defined in this way do not show significant differences:
both lack long range order, either crystal-like or smectic-like.
To that respect, they both look like a liquid, as demonstrated by
the inspection of the density autocorrelation function
(Fig.~\ref{diffrac}). It is difficult, however, to unambiguously
put into evidence the existence of a second-order nonequilibrium
phase transition in our case.

Up to now, we have reported only upon a single point of the
crossover line in the $(T,j)$ plane. A close examination of
Fig.~\ref{lattices} provides some indication about the general
behavior of this line. Indeed, it can be seen that both samples B
and C do not exhibit a strictly homogeneous regime and that the
region close to their bottom boundary tends to be in the fluid
flow regime. This is because, due to edge effects, the average
screening current density tends to be larger in this region. As a
consequence, we may now propose a schematic phase diagram for
flux motion, where we have added the disorder strength as a third
dimension. These observations are not easily compared to the
theoretical expectations for driven lattices (\cite{led98} and
refs therein). First, models investigate the case of a weak
random potential, which is not adequate here, unless the matching
field greatly exceeds the applied magnetic field and $\sigma \ll
1$. Then, the two dimensional, non zero temperature case for the
random potential is far less documented than the three dimensional
one\cite{led98}. We may simply notice that the measured
activation energy at the crossover is close to the plastic energy
barrier (i.e. the barrier experienced by a moving interstitial
defect) $U_{pl} \simeq \varepsilon_{0} /2 \sqrt{3}\pi \simeq
0.4~U_{0}$ (Fig.~\ref{tju})\cite{blatter94} and that it is likely
that this condition holds all along the crossover line. Indeed,
energy barriers larger than $U_{pl}$ cannot be allowed by
decoupled vortex flow channels and larger barriers can be
sustained only by the coupling of these channels in the fluid
flow regime. Finally, we would like to compare with real
relaxation experiments. The main difficulty in realizing the
situation described by our simulations comes from the presence,
in real crystals, of naturally grown point defects. Indeed,
as-grown crystals invariably exhibit at low temperature high
critical current. In the highly anisotropic Bi$_{2212}$ material,
screening current as large as $10^{10} A m^{ -2}$ is observed at
low temperature. The introduction of strong pinning centers, such
as columnar defects, does however induce a sizeable increase of
flux line pinning at low temperature (almost one order of
magnitude for the critical current density at 4.2 K in
Ref.\cite{moshchalkov94}, at a dose $B_{\phi}=20~kG$). From this,
it may be concluded that natural defects are weak or dilute
enough, so as to be overcome by the introduction of extra
columnar defects, and that irradiated crystals might provide a
situation close to the one described here. An expected signature
for the crossover is a weaker temperature dependence of the
(relaxed) screening current in the fluid flow regime - as
compared to the plastic one - due to the increase of the energy
barrier in this regime. A sharp change is indeed observed in the
temperature dependence of the relaxed screening current, for both
as-grown and irradiated samples, although this has been
interpreted as the signature of the crossover between small and
large bundles regime in the first case, and depinning from the
defects in the second one\cite{moshchalkov94}. We believe that a
decisive test should be the measurement of the energy barriers at
this point and the comparison with the one for the plastic
mechanism.

\section{Conclusion}

We have studied the crossover between the fluid flow and the
plastic regimes of the critical state dynamics for a two
dimensional flux lattice, with strong pinning sites corresponding
to a matching field twice the applied field. Following a smooth
trajectory in the $(T,j)$ diagram, we find that the transverse
displacement of the vortex trajectory rises abruptly as this
trajectory crosses the crossover line. There, the apparent
activation energy for flux motion is close to the plastic energy
barrier and increases strongly in the fluid flow regime. We point
out strong similarities between these results and the sharp
crossover observed in the temperature dependence of the screening
current density for as-grown and irradiated Bi$_{2212}$.

%
%
%

\newpage
\newpage
\begin{figure}

\caption{Geometry used for the simulations. The average vortex
flow is along the $x$ direction. Dotted line depicts the periodic
boundary condition.} \label{sketch}
\end{figure}

\begin{figure}

\caption{Top: $(T,j)$ trajectory followed during the simulations.
Bottom: activation energy along the $(T,j)$ trajectory. The
dotted line indicates the plastic barrier, $U_{pl}$. Labels 'B'
and 'C' refer to the samples in Fig.\ref{lattices}.} \label{tju}
\end{figure}

\begin{figure}

\caption{Sample A: static 'critical state' at $T=0$. Lines are
colinear to the local current direction and their length indicate
the current intensity (non linear scale). Samples B and C: same
notation as in Fig.~\ref{tju}; lines are vortices trajectories in
the steady state.} \label{lattices}
\end{figure}

\begin{figure}

\caption{Displacement transverse to the average velocity, along
vortices trajectories, $y(x)$(the average vortex flow is along the
$x$-axis).} \label{diffusion}
\end{figure}

\begin{figure}

\caption{The transverse displacement on the scale of the
equilibrium lattice parameter. Lines are guides to the eye.}
\label{orderparameter}
\end{figure}

\begin{figure}

\caption{Left: in gray scale, density autocorrelation function of
the lattice snapshots. Right: after radial integration. Top and
bottom are respectively sample B and C in Fig.~\ref{diffusion},
both showing liquid order (The exponential decay length, as
defined in Ref.~\cite{bishop91}, is respectively $0.8~a_{0}$ and
$0.5~a_{0}$).} \label{diffrac}
\end{figure}

\begin{figure}

\caption{Schematic phase diagram. The arrow indicates the
trajectory followed in Fig.~\ref{tju}, with indication of samples
B and C in Fig. \ref{tju} and \ref{lattices} .} \label{phases}
\end{figure}
%
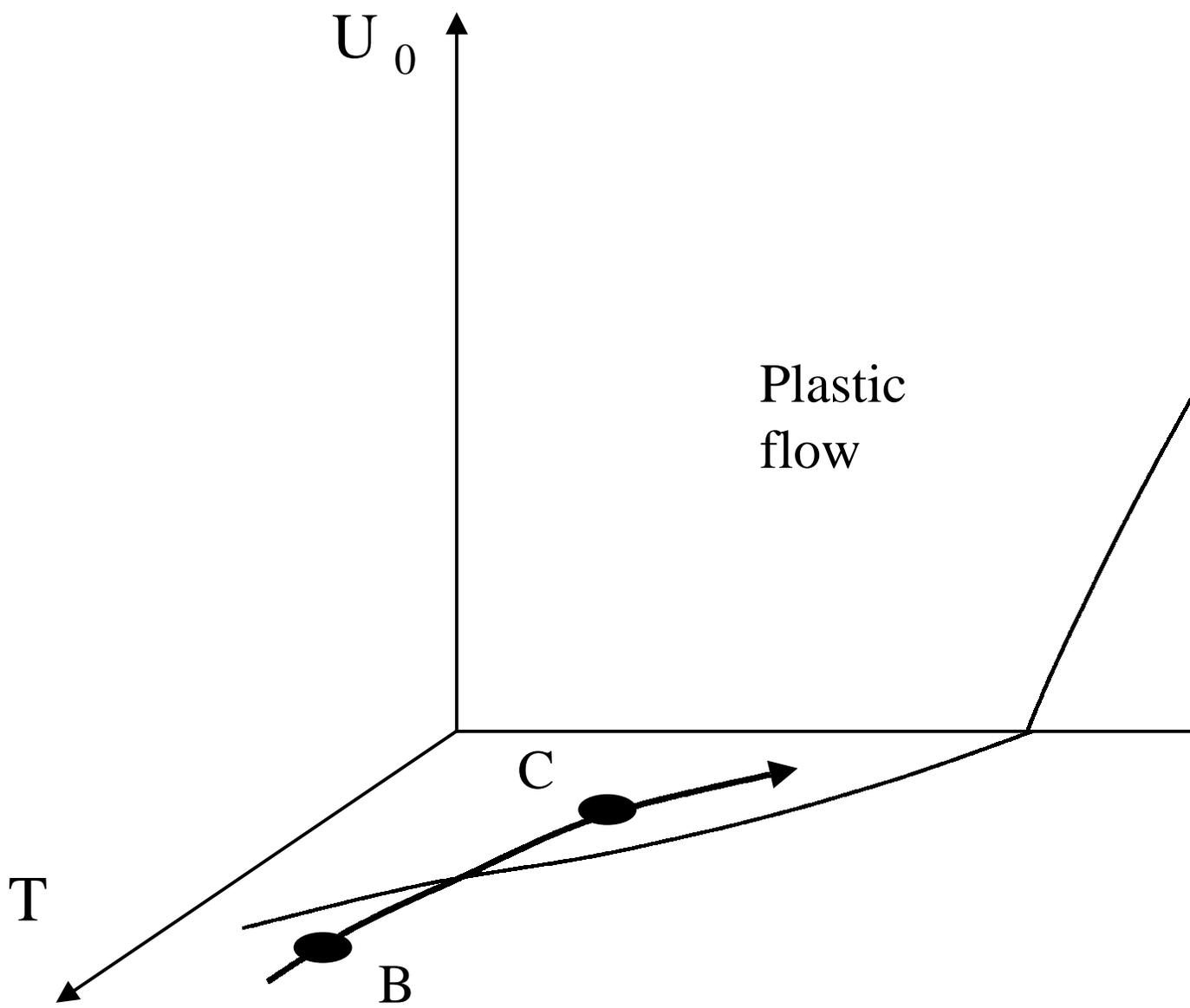
\end{document}